# Compton Spectrum from Poynting Flux Accelerated e+e- Plasma.

Shinya Sugiyama*, Edison Liang*, Koichi Noguchi* and Hideaki Takabe[†]

*Rice University, 6100 Main St. Houston, TX 77005-1892, USA
[†]Institute of Laser Engineering Osaka University, 2-6, Yamada-oka, Suita, OSAKA 565-0871
JAPAN

**Abstract.**
We report the Compton scattering emission from the Poynting flux acceleration of electron-positron plasma simulated by the $2\frac{1}{2}$ dimensional particle-in-cell(PIC) code. We will show these and other remarkable properties of Poynting flux acceleration and Compton spectral output, and discuss the agreement with the observed properties of GRBs and XRFs.

**Keywords:** GRBs, Compton, Plasmas

**PACS:** 98.70.Rz

## INTRODUCTION

We report the Compton scattering emission from the Poynting flux acceleration of electron-positron plasma simulated by the $2\frac{1}{2}$ dimensional particle-in-cell(PIC) code. We study the intense electromagnetic pulse loaded with $e^+e^-$ plasma interacting with ambient blackbody soft photons. Driven by the strong electromagnetic field, the particles can be accelerated to ultrarelativistic energies with a power-law momentum distribution (We call this mechanism the diamagnetic relativistic pulse accelerator (DRPA) which is a special example of Poynting Flux acceleration by comoving transverse EM pulses [1]).

This mechanism results in the development of a power-law spectrum of photons via Compton scattering, (photon index is roughly $-1 \sim -3$). This is consistent with the Gamma-ray, X-ray power-law spectrum from the long duration GRBs. The emission has also strong angle dependence: the emission from 3-8 degree from the Poynting flux direction is harder than from the other angles. We will show these and other remarkable properties of Poynting flux acceleration and Compton spectral output, and discuss the agreement with the observed properties of GRBs and XRFs.

## COMPTON DRAG FORCE

We use a $2\frac{1}{2}$ dimensional explicit simulation scheme based on the PIC method for time advancing of plasma particles and fields [1]. The dimensional system is double periodic in *x* and *z* directions. We solve the Maxwell equations and Lorentz equation of motion

**TABLE 1.** Plasma Parameters

|   | Frequency Ratio $[\Omega_{pe}/\Omega_e]$ | X Length $[\lambda_d]$ | Z Length $[\lambda_d]$ | Plasma Temperature | Thermal Photon Temperature | Photon Density $[n_p/n_e]$ |
|---|---|---|---|---|---|---|
| (A) | 0.1 | 2000 | 4 | 1MeV | 1eV | 1.0 |
| (B) | 0.1 | 2000 | 4 | 1MeV | 1eV | 0.01 |
| (C) | 0.1 | 2000 | 4 | 1MeV | 10eV | 1.0 |
| (D) | 0.01 | 200 | 4 | 1MeV | 1eV | 1.0 |
| (E) | 0.1 | 2000 | 4 | 1MeV | - | - |

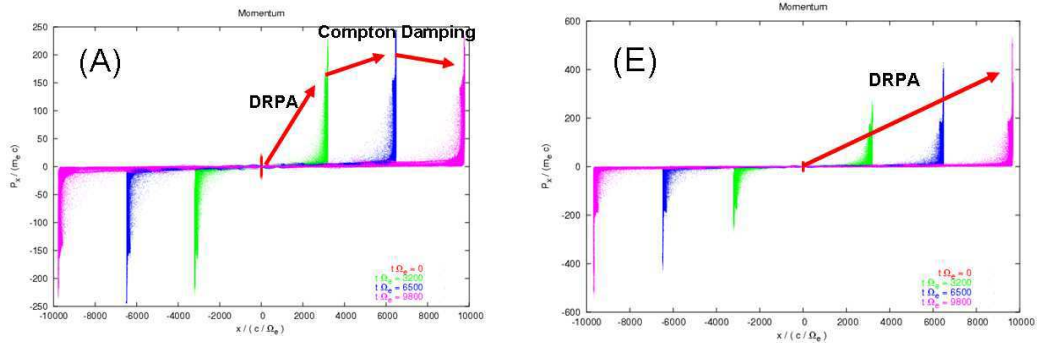

**FIGURE 1.** x-Px Evolution

of the plasma particles including the Compton drag force.

$$m\frac{d(\gamma \vec{v})}{dt} = q\left(\vec{E} + \frac{1}{c}\vec{v} \times \vec{B}\right) - F_\gamma \left(\frac{\vec{v}}{|v|}\right) \quad (1)$$

The last term of the equation ((1)) is the Compton drag force term. $F_\gamma$ means the average photon drag force hitting each particle.

Initially, the electron and positron distributions are assumed to be Maxwellian with uniform temperature $T_{e^-}, T_{e^+}$. The spatial distributions of plasmas have a slab form located in the center of the grid. The initial background magnetic field $\vec{B}_0 = (0, B_0, 0)$ exists only inside the plasma.

## CALCULATION RESULTS

The table(1) is the each initial plasma parameters we calculated. $\Omega_{pe}$ and $\Omega_e$ are the plasma and cyclotron frequency in the initial Maxwellian plasma respectively. $\lambda_d$ is the electron Debye length. $n_p/n_e$ is the number density ratio between photon and electron in a unit length $\lambda_d$.

Case[A] of the table(1) is a higher photon density case. Initially, the particles is accelerated by a DRPA, but, Compton scattering makes the strong drag force to the particle reaching the peak around $\gamma \sim 250$. Case[B] is a lower photon density case. This case is similar to the case[E] which is no photon drag. The case [C] and [D] are the case of higher temperature of the incident photon and strong magnetic field respectively.

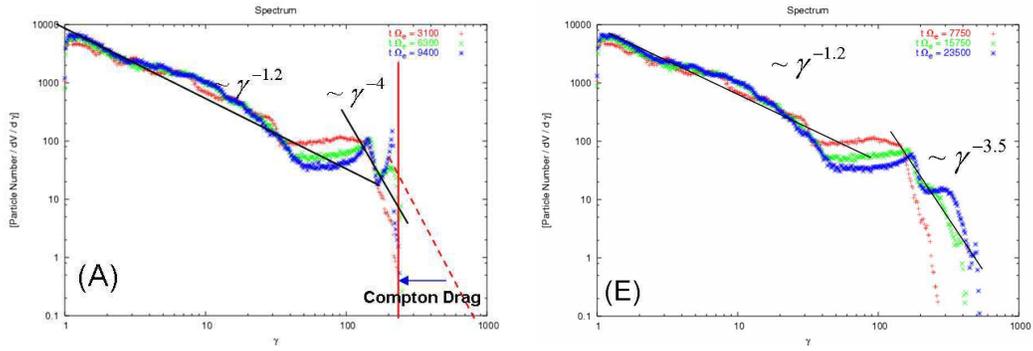

**FIGURE 2.** Gamma Distribution Evolution

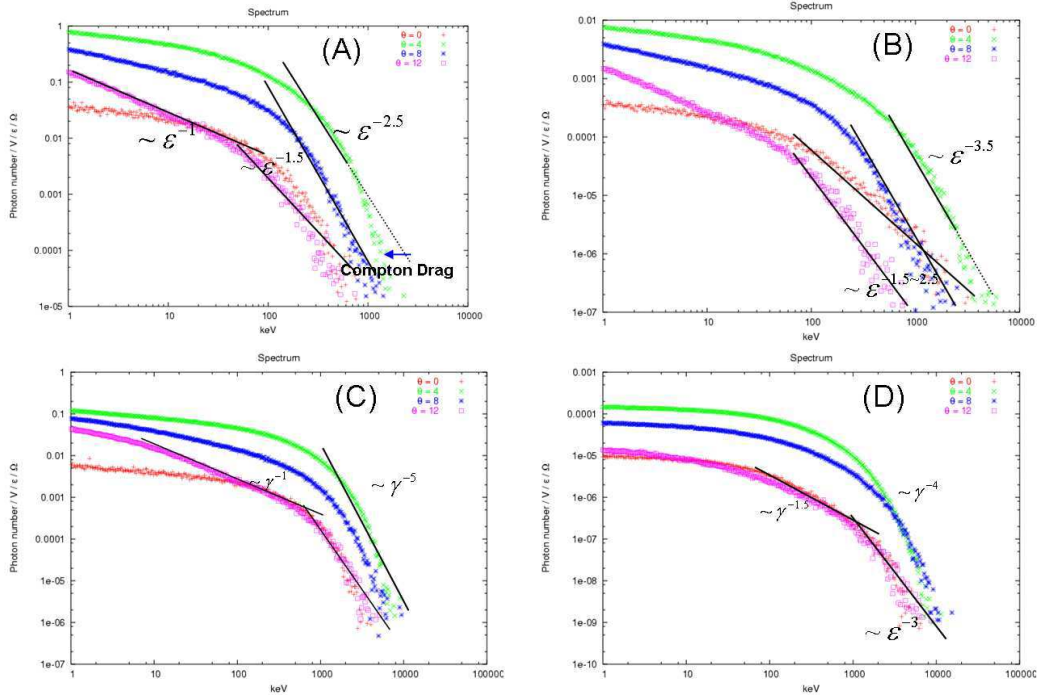

**FIGURE 3.** Angle Dependent Compton Spectrum

Figure (1) shows the global evolution of a DRPA expanding into a vacuum and the Compton drag force effect.

Figure (2) shows the gamma distribution of the particles. The red and green dots on the figure is in the middle of the calculation time, and the blue dots is at the end of the calculation. Case [E] (no Compton effect) shows the strong acceleration, which increases the particle energy to $\gamma \sim 500$ or higher with the time glowing. However, Case [A] shows that the growth of the energy of the particles reach a peak around $\gamma \sim 250$. This means that Compton energy loss is strong and efficient in the case [A].

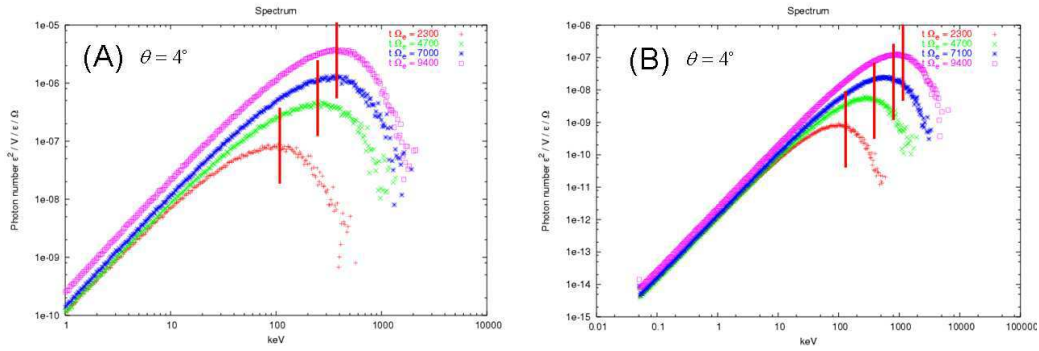

**FIGURE 4.** Spectrum Time Dependence

Figure (3) shows the angle dependent Compton spectrum. In each case, the spectrum around $4° \sim 6°$ has the peak. This is because the main part of DRPA is along the *x* direction, but particles are also accelerated to the vertical direction to the *x* and magnetic direction slightly. Moreover, it seems that there are two power-law indices in lower and higher energy region. The higher energy region of the case[A] has some similarity to the GRB spectrum. Case [A] and [C] have a cut off energy around few *MeV* which is from the strong Compton Drag Force. It is lower efficiency than the Case [B] and [D].

Figure (4) shows the spectrum time dependence along the angle $4°$. In case [A], E-peak is around a few hundred *keV*, this is consistent with the GRB spectrum.

## CONCLUSION

The diamagnetic relativistic pulse accelerator (DRPA) can convert the magnetic energy into the kinetic energy of the surface particles effectively, resulting in ultrarelativistic particles. Most of particle acceleration is in x-direction, but slightly tilted to z-direction, $Pz/Px \sim 0.1 - 0.2$. This causes the Compton spectrum to have strong angle, time dependence.

The particle momentum distribution is roughly $\gamma^{-1.5 \sim 0}$ at low energy, $\gamma^{-3 \sim 4}$ at high energy. This distribution makes the broken power-law Compton spectrum. We can see the power-law Compton spectrum at high energy region of some cases. This power-law and E-peak might be consistent with GRBs or X-ray Flash. However, longer duration simulation will be needed to build up the correct power-law tail.

The Compton drag is effective in the case of high photon density and hot photon surrounding e+e- plasmas. We can see the high energy cut off by the Compton drag in the Compton efficient case because of the Compton drag $\propto \gamma^2$. This cutoff is testable with GRBs, however, The range of E-peak calculated agrees with observed classical GRBs.

## REFERENCES

1. E. Liang and Kazumi Nishimura and Hui Li and S.Peter Gary, *Physical Review Letters* **90**, 085001 (2003).